# Distribution Fitting 2. Pearson-Fisher, Kolmogorov-Smirnov, Anderson-Darling, Wilks-Shapiro, Cramer-von-Misses and Jarque-Bera statistics

## Lorentz JÄNTSCHI[1] and Sorana D. BOLBOACĂ[2]

[1]Technical University of Cluj-Napoca, 400641 Cluj-Napoca; http://lori.academicdirect.org
[2]"Iuliu Hațieganu" University of Medicine and Pharmacy, 400349 Cluj-Napoca, Romania;
sbolboaca@umfcluj.ro

**Abstract.** The methods measuring the departure between observation and the model were reviewed. The following statistics were applied on two experimental data sets: Chi-Squared, Kolmogorov-Smirnov, Anderson-Darling, Wilks-Shapiro, and Jarque-Bera. Both investigated sets proved not to be normal distributed. The Grubbs' test identified one outlier and after its removal the normality of the set of 205 chemical active compounds was accepted. The second data set proved not to have any outliers. Kolmogorov-Smirnov statistic is less affected by the existence of outliers (positive variation expressed as percentage smaller than 2). The outliers bring to Kolmogorov-Smirnov statistic errors of type II and to the Anderson-Darling statistic errors of type I.

## INTRODUCTION

A series of alternatives are available in the literature to measure the agreement between observation and the model. When the model is a distribution law, these alternatives are called statistics. The problem of agreement can be divided in two classes: estimation of (theoretical) distribution parameters and measure of the departure between the model and the observed data. The estimating of the (theoretical) distribution parameters is in details presented in (Jäntschi, 2009).

The present paper presented the common statistics measurement of the departure between observation and the model beside their application on one set of biological active compounds.

## MATERIALS AND METHODS

*Pearson-Fisher Statistic*
Chi Square or Pearson-Fisher ($\chi^2$) test was proposed as a measure of random departure between observation and the theoretical model by Karl PEARSON (Pearson, 1900). The test was later corrected by Ronald FISHER through decrease of the degrees of freedom by a unit (decrease due to the existence of the equality relationship between the sum of observed frequencies and the sum of theoretical frequencies, (Fisher, 1922)), and by the number of unknown parameters of the theoretical distribution when they come as estimated from measures of central tendency (Fisher, 1924).

The agreement between observation and the model are testes though the division of the interval of observation in a given number of intervals (let be *n* this number) by $X^2$ expression given in Eq(1).

$$X^2 = \sum_{i=1}^{n}(O_i - E_i)^2 / E_i \ ; \ p = p_{\chi^2}(X^2, n-t-1) \tag{1}$$

where $X^2$ = chi square statistic; $O_i$ and $E_i$ are the observed and expected frequencies in the *i*-th frequency class; $p_{\chi 2}$ = probability of observing the $X^2$ departure (from 0) by using chi square distribution; $n$ = number of classes; $t$ is the number of central tendency measures estimated parameters).

Usually the agreement is accepted when *p* is no less than 5%. Despite the fact that $\chi^2$ test is the best known statistic for testing the agreement between observation and the model, the frame of the application of it is one of the complex ones (Snedecor and Cochran, 1989).

A first issue is about the number of the frequency classes; a series of options are available, from which two of them are as follows:

÷ Rounding the Hartley entropy (Hartley, 1928) of observation: $log_2(2N)$, where *N* is the number of observations (EasyFit software[1] uses this approach);
÷ Obtaining simultaneously of classes number and classes width from the histogram seen as an estimator of the density (Scott, 1992) when an optimal criteria is chosen from a list (Dataplot[2] automatically generate the frequency classes using a rule of this type: width of a frequency class is 0.3·*s*, where *s* is the sample standard deviation; highest class and lowest class are given by a departure of ±6·*s* relative to the sample mean and the marginal classes with zero observed frequency are omitted).

A second issue is about the width of the frequency classes. At least two approaches are available: the data can be grouped in equal (theoretical or observed) probability or the data can be grouped in equal (theoretical or observed) width. First approach (equal probability) is more often used due to its better efficiency for grouped data.

Other issue take into consideration the number of observations in a frequency class. In order to keep only the random effect, at least 5 observations are required in every class. Thus, in practice, when a procedure give the frequency classes with less than 5 observations per class, adjacent classes are joined together to fulfil this requirement.

### *Kolmogorov-Smirnov Statistic*

Kolmogorov-Smirnov statistic uses comparison of the cumulative frequencies. The Kolmogorov-Smirnov statistic may act as measure of the agreement between observation and the model (Kolmogorov, 1941) as well as a measure between two series of observations (Smirnov, 1948). If $F_t(X_i)$ and $F_o(X_i)$ are the theoretical and the observed cumulative frequencies for distinct and ascending ordered observations $X_i$, then a series of statistics are based on $F_t(X_i)$-$F_o(X_i)$ difference:

$$D_- = \max_{1 \le i \le n}(F_t(X_i) - F_o(X_i)), \ D_+ = \max_{1 \le i \le n}(F_o(X_i) - F_t(X_i))$$
$$D = \max(D_-, D_+), \ V = D_- + D_+ \tag{2}$$

where $D_-$ measures the largest leak in observation compared with theoretical value, $D_+$ measures the largest excess in observation compared with theoretical value, D measures the largest difference in between (it is the most frequent used), and V represent the modification of D given by Kuiper (Kuiper, 1960).

For the $K = D\sqrt{N}$ statistic the distribution law comes from:

$K = \max_{0 \le t \le 1}|B(t)|$, where B is the Brownian bridge conditioned by:

---

[1] http://www.mathwave.com
[2] http://www.itl.nist.gov/div898/software/dataplot.html

$B(0) = B(1) = 0$ (boundaries), $M(B(t)) = 0$ (mean), $Var(B(t)) = t(t-1)$ (variance) and the probability of observing a larger departure ($K < x$) is given by:

$$P(K<x) = 1 - 2\sum_{i=1}^{\infty}(-1)^{i-1}e^{-2i^2K^2} = \frac{\sqrt{2\pi}}{K}\sum_{i=1}^{\infty}e^{-(2i-1)^2\pi^2/8K^2} \qquad (3)$$

Similarly, for V statistic:

$$P(V<x) = Q_{KP}(\sqrt{n}+0.155+0.24/\sqrt{n}); \quad Q_{KP}(\lambda) = 2\sum_{i=1}^{\infty}(4i^2\lambda^2-1)e^{-2i^2\lambda^2} \qquad (4)$$

*Anderson-Darling Statistic*

The Anderson-Darling statistic (Anderson and Darling, 1952) uses the distinct values from the observed data ordered ascending. The $A^2$ statistic is computed as follows:

$$A^2 = -N - \sum_{k=1}^{N}\frac{2k-1}{N}\left(\ln(F(X_k)) + \ln(1-F(X_{n+1-k}))\right) \qquad (5)$$

and has the variance given by:

$$Var(A^2) = \frac{2(\pi^2-9)}{3} + \frac{10-\pi^2}{N} \qquad (6)$$

The probability of the worst observation depends on the theoretical distribution. For the normal distribution this probability is given by (Trujillo-Ortiz et al., 2007):

$$A^2_c = A^2(1+0.75/n+2.25/n^2)$$

$$p = \begin{cases} 1 - \exp(-13.436+101.14\ A_c - 223.73\ A_c^2), & A_c < 0.2 \\ 1 - \exp(-8.318+42.796\ A_c - 59.938\ A_c^2), & 0.2 \leq A_c < 0.34 \\ \exp(0.9177 - 4.279\ A_c - 1.38\ A_c^2), & 0.34 \leq A_c < 0.6 \\ \exp(1.2937 - 5.709\ A_c + 0.0186\ A_c^2), & A_c \leq 0.6 \end{cases} \qquad (7)$$

*Wilks-Shapiro Statistic*

Wilks-Shapiro statistic (Shapiro and Wilk, 1965) measure a departure from normality. The $W^2$ statistic is computed on ordered values of observed data as follows:

$$W^2 = \left(\sum_{i=1}^{N}A_iX_i\right)^2 \Big/ \sum_{i=1}^{N}(X_i-\overline{X})^2, \quad A = m^TV^{-1}(m^TV^{-1}V^{-1}m^T)^{-1/2} \qquad (8)$$

where *m* and *V* = expected values of the order statistics of independent and identically-distributed random variables sampled from the standard normal distribution and their covariances respectively. The agreement is rejected if $W^2$ value is too high. The expression of critical values as well as exact probabilities associated with the observed values can be obtained via software (Royston, 1995).

*Cramer-von-Misses Statistic*

Then the Cramér-von-Mises statistic (CMS) is:

$$W^2 = \frac{1}{n}\left(\frac{1}{12n} + \sum_{i=1}^{n}\left(\frac{2i-1}{n}-F(X_i)\right)^2\right) \qquad (9)$$

The probability to observe a value greater than CMS is given by the following approximate function (Levin, 2003):

$$p(W^2<x) = 0.67\cdot\exp(-5.6\cdot W^2) \qquad (10)$$

*Jarque-Bera Statistic*

The Jaque-Bera statistic (Jarque and Bera, 1981) uses third and fourth central moments as measures of the departure between observed distribution and a normal distribution. The statistic comes from:

$$JB = N\frac{g_1^2 + g_2^2/4}{6}; \quad g_1 = \frac{m_3}{m_2^{3/2}}; \quad g_2 = \frac{m_4}{m_2^2} - 3; \quad m_k = \frac{\sum_{i=1}^{N}(X_i - \overline{X})^k}{N} \quad (11)$$

and the probability of observation comes from Chi Square distribution with 2 degrees of freedom ($p = p_{\chi^2}(JB, 2)$).

*Z-Based Statistics*

For central tendency measures, the normal distribution assumption operates as well on it. The following statistics are available:

$$z_{mean} = \frac{\overline{X} - \mu}{s}\sqrt{N}; \quad z_{variance} = \frac{(N-1)s^2 - \sigma^2 N}{\sqrt{(N-1)^3 m_4 - (N-3)m_2^2}} N$$

$$z_{st.dev} = \frac{c_4 \frac{s}{\sigma} - 1}{\sqrt{1 - c_4^2}}; \quad c_4 = \frac{\Gamma(n/2)\sqrt{\frac{2}{n-1}}}{\Gamma((n-1)/2)} \quad (12)$$

$$z_{skewness} = g_1\sqrt{\frac{(N+1)(N+3)}{6(N-2)}}; \quad z_{kurtosis} = \left(\frac{N+1}{N-1}g_2 - 3\right)\sqrt{\frac{(N+3)(N+5)(N-1)^2}{24N(N-2)(N-3)}}$$

*Grubbs' Statistic*

Many statistical techniques are sensitive to the presence of outliers, Grubbs being one of them. All calculations, including the mean and standard deviation may be distorted by a single grossly inaccurate data point. Checking for outliers should be a routine part of any data analysis. Grubbs' test is used to detect outliers in a univariate data set (Grubbs, 1969). It is based on the assumption of normality. The Grubbs' test statistic is the largest absolute deviation from the sample mean ($\overline{X}$) in units of the sample standard deviation (s). The Grubbs test statistics may be applied to the minimum and to the maximum (Eq(13)) and to the both (Eq(14) when associated probabilities of observed are obtained from Student t distribution:

$$G_{min} = \frac{\overline{X} - \min(X)}{s}, \quad G_{max} = \frac{\max(X) - \overline{X}}{s}$$

$$p_G = 2n \cdot p_t\left(G\frac{\sqrt{n(n-2)}}{n-1}, n-2, 1\right) \quad (13)$$

$$G_{all} = \frac{\max(\overline{X} - \min(X), \max(X) - \overline{X})}{s},$$

$$p_G = n \cdot p_t\left(G\frac{\sqrt{n(n-2)}}{n-1}, n-2, 2\right) \quad (14)$$

Under assumption of normal distributed errors the Wilk-Shapiro, Jarque-Bera, and Grubbs' statistics can be applied for detecting outliers, obtaining thus powerful tests for detecting outliers for any model of observed data.

*Applications*

Two sets of investigated data were taken from literature in order to illustrate the procedures described above. First set (Duchowicz et al., 2008) records the measurements of aqueous solubility (Sol) for a series of 166 drug-like compounds. The second set (Jäntschi et al., 2009) the measurements of octanol-water partition coefficient ($K_{ow}$) for a series of 206 polychlorinated biphenils expressed both in logarithmic scale ($\log_{10}(Sol)$, $[Sol]=mg \cdot ml^{-1}$; $\log_{10}(K_{ow})$, $[K_{ow}]=1$). Table 1 contain the experimental values for each set in ascending order.

The following tests were applied on the investigated experimental data: Kolmogorov-Smirnov (abbreviated as KS), Anderson-Darling (AD), Chi Square (CS), Wilks-Shapiro (WS), Z-based Skewnes ($Z_{Skewness}$), Z-based Kurtosis ($Z_{Kurtosis}$), Jarque-Bera (JB).

Tab. 1.
Two data sets of measurements under assumption of normal distribution

| log(Sol) for 166 drug-like compounds | log($K_{ow}$) for 206 polychlorinated biphenils |
|---|---|
| -6; -5.53; -5.376; -4.247; -4.173; -4; -3.699; -3.61; -3.522; -3.397; -3.239; -3.125; -3.096; -3.07; -3.046; -3; -2.85; -2.795; -2.747; -2.656; -2.397; -2.318; -2.221; -2.136; -2.113; -2.041; -2; -2; -2; -1.958; -1.886; -1.853; -1.853; -1.826; -1.823; -1.823; -1.728; -1.673; -1.657; -1.61; -1.588; -1.397; -1.397; -1.392; -1.337; -1.301; -1.301; -1.252; -1.23; -1.216; -1.209; -1.124; -1.051; -1.045; -1.003; -0.951; -0.854; -0.854; -0.841; -0.821; -0.767; -0.701; -0.699; -0.658; -0.652; -0.648; -0.62; -0.602; -0.585; -0.523; -0.523; -0.495; -0.495; -0.495; -0.444; -0.426; -0.397; -0.367; -0.301; -0.225; -0.222; -0.22; -0.187; -0.161; -0.125; -0.102; -0.08; -0.056; -0.051; -0.009; 0; 0; 0.013; 0.017; 0.026; 0.057; 0.077; 0.079; 0.079; 0.079; 0.079; 0.146; 0.204; 0.258; 0.301; 0.301; 0.301; 0.398; 0.431; 0.477; 0.602; 0.623; 0.633; 0.663; 0.699; 0.699; 0.699; 0.7; 0.769; 0.78; 0.806; 0.826; 0.845; 0.846; 0.873; 0.886; 0.912; 0.933; 1.012; 1.053; 1.079; 1.079; 1.103; 1.187; 1.294; 1.301; 1.397; 1.414; 1.519; 1.536; 1.556; 1.623; 1.658; 1.698; 1.698; 1.698; 1.763; 1.857; 1.872; 1.914; 1.929; 1.934; 2.146; 2.214; 2.221; 2.301; 2.396; 2.522; 2.557; 2.665; 2.698; 2.698; 2.698; 2.77; 2.806; 3.352 | 4.151; 4.401; 4.421; 4.601; 4.941; 5.021; 5.023; 5.15; 5.18; 5.295; 5.301; 5.311; 5.311; 5.335; 5.343; 5.404; 5.421; 5.447; 5.452; 5.452; 5.481; 5.504; 5.517; 5.537; 5.537; 5.551; 5.561; 5.572; 5.577; 5.577; 5.627; 5.637; 5.637; 5.667; 5.667; 5.671; 5.677; 5.677; 5.691; 5.717; 5.743; 5.751; 5.757; 5.761; 5.767; 5.767; 5.787; 5.811; 5.817; 5.827; 5.867; 5.897; 5.897; 5.904; 5.943; 5.957; 5.957; 5.987; 6.041; 6.047; 6.047; 6.047; 6.057; 6.077; 6.091; 6.111; 6.117; 6.117; 6.137; 6.137; 6.137; 6.137; 6.137; 6.142; 6.167; 6.177; 6.177; 6.177; 6.204; 6.207; 6.221; 6.227; 6.227; 6.231; 6.237; 6.257; 6.267; 6.267; 6.267; 6.291; 6.304; 6.327; 6.357; 6.357; 6.367; 6.367; 6.371; 6.427; 6.457; 6.467; 6.487; 6.497; 6.511; 6.517; 6.517; 6.523; 6.532; 6.547; 6.583; 6.587; 6.587; 6.587; 6.607; 6.611; 6.647; 6.647; 6.647; 6.647; 6.647; 6.657; 6.657; 6.671; 6.671; 6.677; 6.677; 6.677; 6.697; 6.704; 6.717; 6.717; 6.737; 6.737; 6.737; 6.747; 6.767; 6.767; 6.767; 6.797; 6.827; 6.857; 6.867; 6.897; 6.897; 6.937; 6.937; 6.957; 6.961; 6.997; 7.027; 7.027; 7.027; 7.057; 7.071; 7.087; 7.087; 7.117; 7.117; 7.117; 7.121; 7.123; 7.147; 7.151; 7.177; 7.177; 7.187; 7.187; 7.207; 7.207; 7.207; 7.211; 7.247; 7.247; 7.277; 7.277; 7.277; 7.281; 7.304; 7.307; 7.307; 7.321; 7.337; 7.367; 7.391; 7.427; 7.441; 7.467; 7.516; 7.527; 7.527; 7.557; 7.567; 7.592; 7.627; 7.627; 7.657; 7.657; 7.717; 7.747; 7.751; 7.933; 8.007; 8.164; 8.423; 8.683; 9.143; 9.603 |

RESULTS AND DISCUSSION

The distributions of the investigated experimental data expressed graphically by using histogram plot are presented in Figure 1. The analysis of the Duchowicz et al. data set

(Duchowicz et al., 2008) seems to be more skewed compared to the Jäntschi et al. data set (Jäntschi et al, 2009).

Seven tests were applied in order to measurement of the departure between observation and the model. The results are presented in Tables 2 and 3.

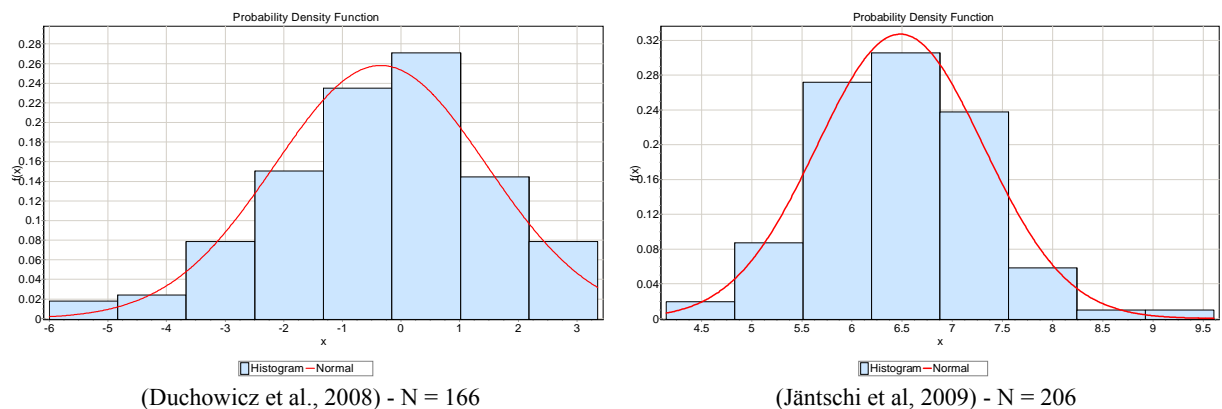

(Duchowicz et al., 2008) - N = 166        (Jäntschi et al, 2009) - N = 206
Fig. 1. Histograms of observed biological activities

Tab. 2.
Hypothesis of normality: experimental biologic activity for Duchowicz et al. set (Duchowicz et al., 2008)

| Statistic | Value | Probability of observation | Reject the hypothesis of normality |
|---|---|---|---|
| Kolmogorov-Smirnov | 0.05508 | 67.43% | No |
| Anderson-Darling | 0.56539 | 14.1%; 12.5%; 14.3% | No |
| Chi Squared | 3(df=7) | 88.6% | No |
| Wilks-Shapiro | 0.98173 | 2.8% | Yes |
| $Z_{Skewness}$ | -2.58 | 1‰ | Yes |
| $Z_{Kurtosis}$ | 0.53 | 59.5% | No |
| Jarque-Bera | 6.61 | 3.7% | Yes |

Tab. 3.
Hypothesis of normality: experimental biologic activity for Jäntschi et al. set (Jäntschi et al., 2009)

| Statistic | Value | Probability of observation | Reject the hypothesis of normality |
|---|---|---|---|
| Kolmogorov-Smirnov | 0.03348 | 96.91% | No |
| Anderson-Darling | 0.44432 | 27.2%; 25.2%; 19.2% | No |
| Chi Squared | 11(df=7) | 13.8% | No |
| Wilks-Shapiro | 0.98709 | 5.8% | No |
| $Z_{Skewness}$ | 1.48 | 14% | No |
| $Z_{Kurtosis}$ | 2.51 | 1.2% | Yes |
| Jarque-Bera | 7.577 | 2.3% | Yes |

The Jarque-Bera test is not affected by the tied values while the Kolmogorov-Smirnov and Anderson-Darling test are affected as results from Table 2.

The analysis of the results presented in table 2 and 3 revealed that the hypothesis of normality do not be accepted for the investigated data sets. This hypothesis is rejected by 3 tests when Duchowicz et al. set (Duchowicz et al., 2008) is analyzed and by 2 tests when

Jäntschi et al. (Jäntschi et al., 2009) set is investigated. These results lead to the conclusion that both sets contain outliers.

The Grubbs' test was applied in order to identify the outliers. The Grubbs' test did not identify any outlier for the Duchowicz et al. set. The following results were identified for the Jäntschi et al. set for a significance level of 5% (α=5%):
- One outlier: experimental data of 9.603 (5%).
- No outlier (5%): Y\{9.603}.

The outlier identified by Grubb's test was removed from the Jäntschi et al. set and the same seven tests were again applied in order to measurement of the departure between observation and the model. The results are presented in Table 4.

Tab. 4.
Hypothesis of normality: experimental biologic activity for Jäntschi et al. set after remodel of the identified outlier (μ = 6.4653; σ = 0.80344)

| Statistic | Value | Probability of observation | Reject the hypothesis of normality |
|---|---|---|---|
| Kolmogorov-Smirnov | 0.03579 | 94.68% | No |
| Anderson-Darling | 0.37878 | 40.3%; 39.5%; 21.0% | No |
| Chi Squared | 8.64(df=7) | 27.9% | No |
| Wilks-Shapiro | 0.98709 | 47.8% | No |
| $Z_{Skewness}$ | 1.48 | 79.2% | No |
| $Z_{Kurtosis}$ | 2.51 | 41.5% | No |
| Jarque-Bera | 0.56146 | 75.5% | No |

## CONCLUSIONS

Kolmogorov-Smirnov statistic is less affected by the existence of outliers (positive variation expressed as percentage smaller than 2). The outliers bring to Kolmogorov-Smirnov statistic errors of type II (the null hypothesis is accepted even if it is not true). The Kolmogorov-Smirnov statistic is followed by Anderson-Darling (variation of 10%). The outliers bring to Anderson-Darling statistic errors of type I (the null hypothesis is rejected even if it is true). The other statistics are more affected by the existence of outliers in the following order: Chi-Square (100%), $Z_{Skewness}$ (470%), Jarque-Bera (3200%), $Z_{Kurtosis}$ (3400%).

*Acknowledgments*. The research was partly supported by UEFISCSU Romania through ID1051/202/2007 grant.